# Quantifying and analyzing rock trait distributions of rocky fault scarps using deep learning

Zhiang Chen[1,*], Chelsea Scott[1], Devin Keating[1], Amanda Clarke[1], Jnaneshwar Das[1], Ramon Arrowsmith[1]

[1] School of Earth and Space Exploration, Arizona State University, Tempe, AZ 85281, USA; {cpscott1, dkeatin2, Amanda.Clarke, jnaneshwar.das, ramon.arrowsmith}@asu.edu
[*] Corresponding: zch@asu.edu

## Abstract

We apply a deep learning model to segment and identify rock characteristics based on a Structure-from-Motion orthomap and digital elevation model of a rocky fault scarp in the Volcanic Tablelands, eastern California, in the United States. By post-processing the deep learning results, we build a semantic rock map and analyze rock trait distributions. The resulting semantic map contains nearly 230,000 rocks with effective diameters ranging from 2 cm to 250 cm. Rock trait distributions provide a new perspective on rocky fault scarp development and extend past research on scarp geometry including slope, height, and length. Heatmaps indicate rock size spatial distributions on the fault scarp and surrounding topographic flats. Median grain size changes perpendicular to the fault scarp trace with the largest rocks on the downslope proximal to the scarp footwall. Correlation analyses illustrate the relationship between rock trait statistics and fault scarp geomorphology. Local fault scarp height correlates with median grain size ($R^2$ of 0.6), the mean grain size of the largest rocks ($R^2$ of 0.8), and the ratio of the number of small to large rocks ($R^2$ of 0.4). The positive correlation ($R^2$ of 0.8) between local fault scarp height and standard deviation of grain size suggests that rocks on a higher fault scarp are less well sorted. The correlation analysis between fault scarp height and rock orientation statistics supports a particle transportation model in which locally higher fault scarps have relatively more rocks with long axes parallel to fault scarp trace because rocks have a larger distance to roll and orient the long axes. Our work demonstrates a data-driven approach to geomorphology based on rock trait distributions, promising a greater understanding of fault scarp formation and tectonic activity, as well as many other applications for which granulometry is an indicator of process.

# 1. INTRODUCTION

Fault scarps serve as surface expressions of tectonic faulting and have geomorphic characteristics that indicate the style and timing of tectonic activity (Wallace, 1977; Stewart & Hancock, 1990; Figueiredo & Nash, 2022). For example, the maximum angle of a typical alluvial fault scarp's slope increases with scarp height and decreases with scarp age (Bucknam & Anderson, 1979). Lateral changes in fault scarp morphology can indicate the fault's segmentation (e.g., Crone & Haller, 1991). Earthquake magnitude positively correlates with fault displacement and surface rupture length (Bonilla et al., 1984). These geomorphic characteristics (e.g., scarp slope, height, and length) were multimeter-scale features. Those early studies also focused largely on scarps developed in poorly consolidated alluvial fan sediments. In this paper, we investigate rocky fault scarps in the Volcanic Tablelands of eastern California (USA) where faulted rock masses comprising the fault scarps are large enough to be readily observed (Figures 1-3; Ferrill et al., 2016). We examine geometric characteristics of grains (rock traits) on the scarp face, such as size, aspect ratio, and orientation. We establish a method to determine simple rock traits at scale using deep learning and demonstrate that these traits have the potential to characterize rocky fault scarp formation processes.

We applied a generalizable approach to rock trait extraction with consideration of the challenges associated with rocky fault scarps. With the extracted rock traits, we analyzed their relationship with fault scarp geomorphology to explore the relative roles of three major processes of rocky fault scarp formation (i.e., columnar jointing, tectonic fracturing, and geomorphic fracturing). Specifically, for the rock trait extraction approach, we adopted a pipeline of UAS, SfM, and deep learning (UAS-SfM-DL) from recent work (Chen et al., 2020), which segmented individual rocks based on the combined information from both an RGB orthomap and a DEM. To solve the rock splitting problem, we applied an instance registration algorithm that merged rocks on the sliced tile edges. Post-processing reduced deep learning false-negative segmentation and extracted rock traits. From the post-processing results, we investigated rock trait distributions over both the entire study area and along the segmented fault scarp within it. We also conducted correlation analyses between rock trait statistics and rocky fault scarp geomorphology, which demonstrated a new perspective for studying rocky fault scarp formation.

# 2. RELATED WORK AND BACKGROUND

## 2.1. Rocky Fault Scarps

Classical fault scarp geomorphic analyses have been summarized by Hanks et al. (1984) and Hanks (2000). In their work on "alluvial scarps", the processes of the fault scarp formation are assumed to be transport-limited. Scarp profile development was idealized as diffusive such that elevation change over time scales with scarp curvature and a generalized transport rate term (Culling, 1960). Andrews and Bucknam (1987) and many others subsequently deployed a non-linear slope dependence for transport rate. Morphologic modeling of profile development and morphologic dating of fault scarps using calibrated transport rates have been widely deployed (e.g., Nash, 1980; Avouac, 1993; Arrowsmith, et al., 1998; Hanks, 2000; Xu, et al., 2021; and many others). Scarps consisting of solid bedrock exposing the grooved fault surface have been the target of cosmogenic radionuclide dating (e.g., Benedetti et al., 2002; Schlagenhauf, et al., 2010; and numerous others). The geomorphic development of these bedrock scarps is production-limited: transport cannot occur until enough material has been produced by weathering processes (e.g., Anderson & Humphrey, 1989; Arrowsmith et al., 1996, 1998). In this paper, we examine an intermediate feature between transport-limited and production-limited scarps: rocky fault scarps. For them, the faulted rock masses and particles comprising the fault scarps are large enough to be readily observed. And the particle transport is not easily idealized as diffusive, nor is it purely production-limited.

Rock trait statistics provide a new perspective to understand rocky fault scarp formation. In faulted rock masses such as those we examined in the Volcanic Tablelands, the formation of a rocky normal fault scarp involves three major processes in the following sequence: columnar jointing associated with cooling of the volcanic unit (or jointing from distributed fracturing in hard layered rocks), localized tectonic faulting, and geomorphic fracturing (including transport of blocks exposed on or near the scarp face), as depicted in Figure 1. The tectonic and geomorphic fracturing may overlap in time, but the general progression holds. Each process contributes a distribution of particle sizes with progressive reduction of size with increasing fracturing.

Initial cooling of extrusive volcanic rocks (welded pyroclastic flows in the case of Bishop Tuff) forms polygonal fracture sets (Figures 1 and 2; *columnar jointing*; see review by Hetényi, et al., 2012). The columns form as their bounding fractures propagate downward, following the cooling front. Slower cooling and rhyolitic compositions promote larger columnar joint spacing with bounding polygonal joint segment lengths as large as 3-4 m (Hetényi, et al., 2012). Secondary, horizontal fracturing segments the columns into blocks (e.g., Yang and Wu, 2006), which are then weathered physically and chemically to produce smaller fragments and round the blocks (e.g., Moon and Jayawardane, 2004). In the case of the Bishop Tuff, this set of blocks is exposed now (typical diameters as much as 1-2 m) on the tops of the scarps, and from there blocks are further fractured and transported (see Figure 3 in Ferrill, et al., 2016; Figures 1 and 2).

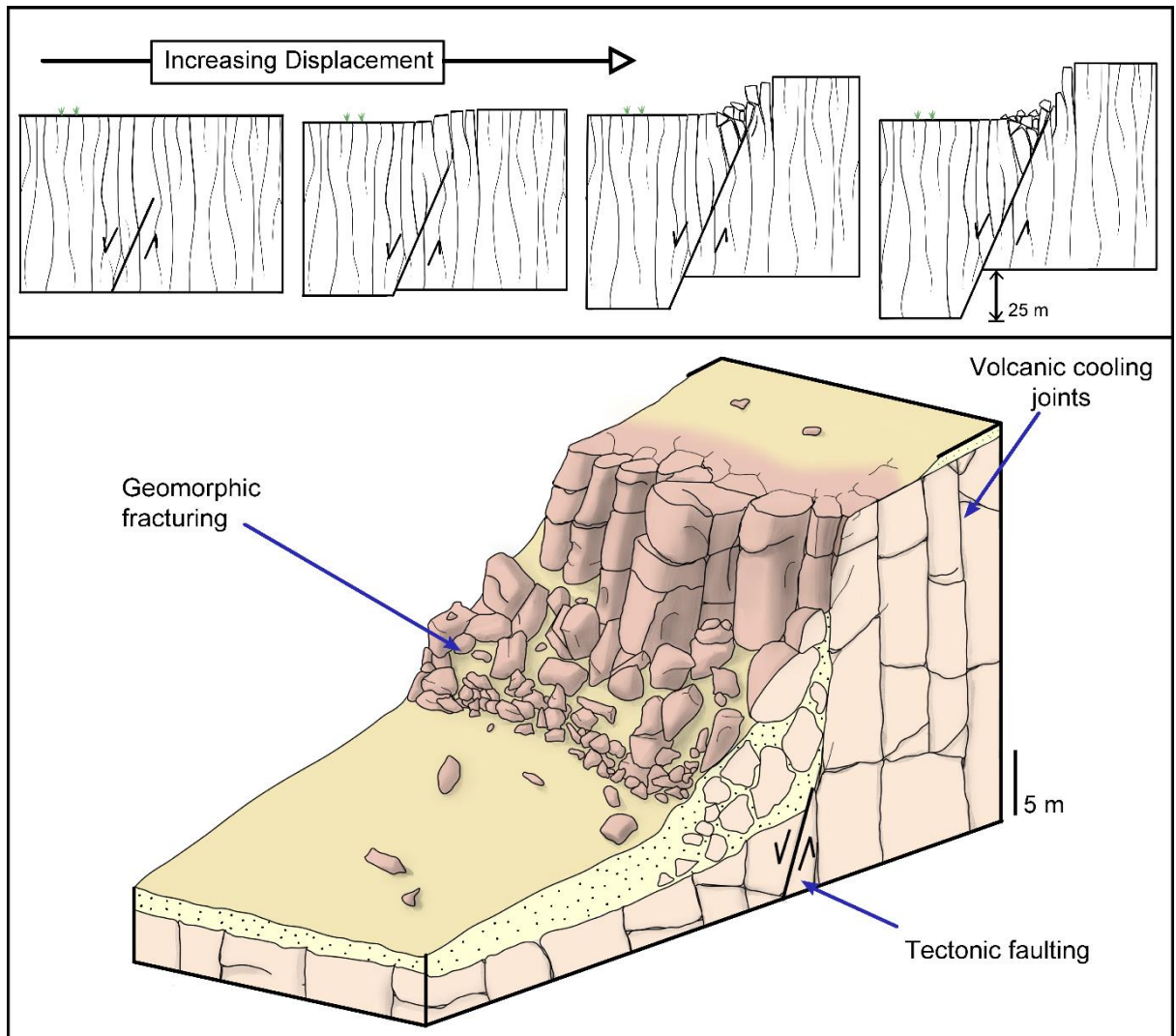

Figure 1. Formation of rocky fault scarps in the Volcanic Tablelands involves three major processes: columnar jointing, tectonic faulting, and geomorphic fracturing. Columnar jointing forms large, polygonal rocks initially. Along with erosion, tectonic faulting breaks, shears, rotates, and crushes the rocks along the fault zone and exposes the rock mass interior. Geomorphic fracturing results from toppled rocks breaking apart into smaller pieces from collisions and cyclic thermal stresses.

Faulting of the columnar-jointed Bishop Tuff has been the focus of many studies (see review by Ferrill, et al., 2016). As the fault offset accumulates, a monocline may form and then breach as the deformation localizes (Grant and Kattenhorn, 2004). The fault offset exposes the fractured columns (Ferrill et al., 2016; Blakeslee & Kattenhorn, 2013). The columnar jointing-bounded blocks rotate, and they may be broken and crushed (Ferrill, et al., 2016). We refer to this process here as *tectonic fracturing*.

Grain size reduction due to faulting depends on volumetric and distortional deformation, confining stress, and dynamic stresses during the passage of earthquake ruptures (e.g., Sammis and Ben Zion, 2008;

Griffith, et al., 2018). Dilation and rotation appear to dominate the faulting-related deformation associated with the ~20 m of throw along the fault examined here. Modest crushing may occur, but confining stresses are low. We do not know if these faults have experienced dynamic rupture with high slip speeds or if they are more sympathetic to deeper coseismic slips.

With further offset and the action of surface processes, the blocks may topple, roll and slide. Impacts such as thermal cycling and fatigue (e.g., Eppes, et al., 2010; Eppes & Keanini, 2017) along with physical and chemical weathering of the particle surfaces (e.g., Moon and Jayawardane, 2004) drive further fracturing and particle diminution. We refer to this grain size reduction process as *geomorphic fracturing*.

Here, we examine general correlations between rock trait distributions and fault scarp geomorphology, which will help to indicate the relative roles of these processes (i.e., columnar jointing, tectonic fracturing, and geomorphic fracturing) in the development of rocky scarps. We are not able to quantitatively deconvolve the relative contributions of the fracturing by volcanic, tectonic, and geomorphic processes, but rather qualitatively recognize their contributions to the observed particle size distributions. This example also serves as a demonstration of our approach and its potential for broad applications across geomorphology, structural geology, and volcanology in particular.

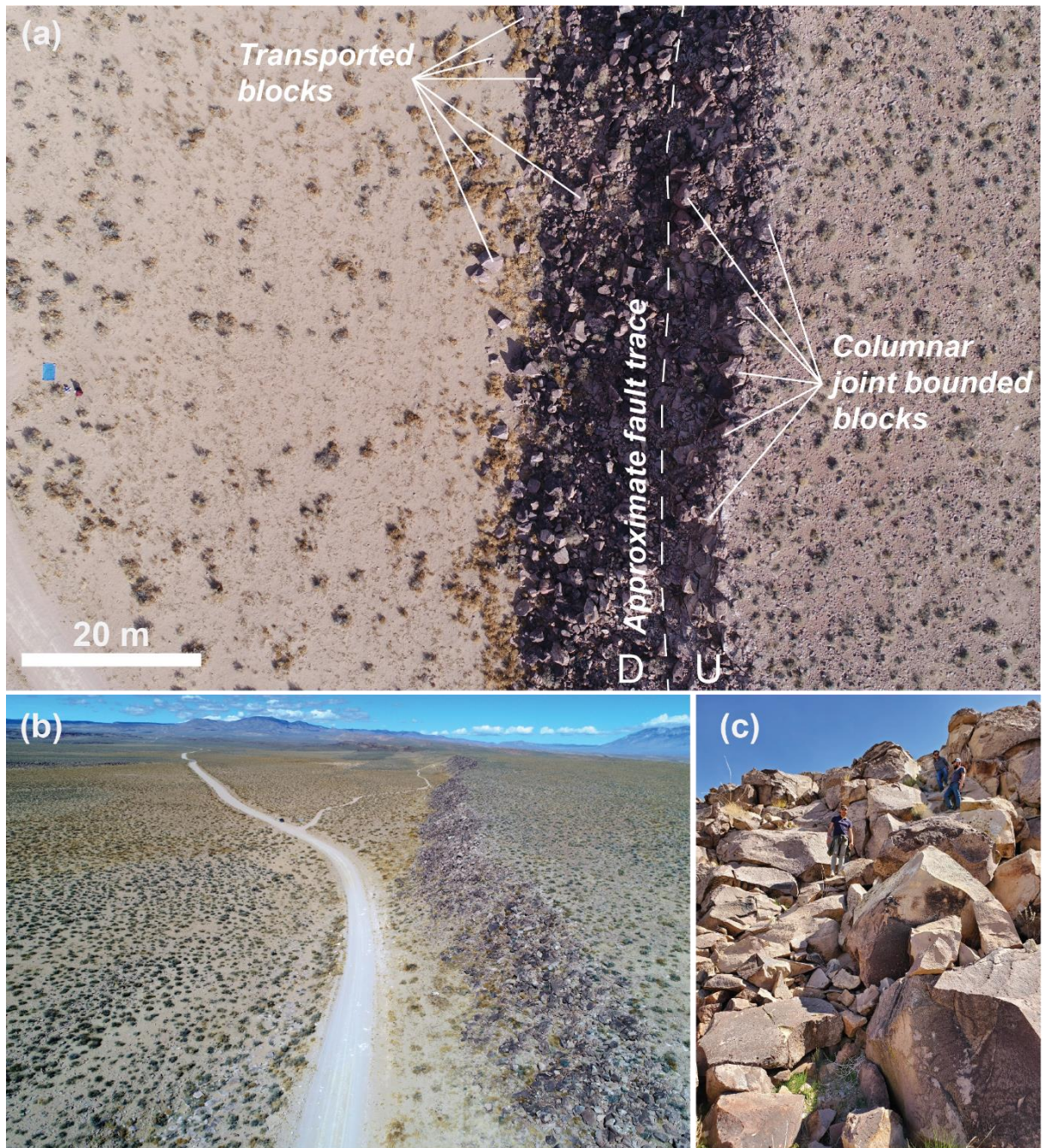

Figure 2. Rocky normal fault scarp in Volcanic Tablelands, eastern California. (a) Vertical view illustrating example columnar-joint bounded columns, the approximate fault trace (Up and Down-thrown sides indicated), and example transported blocks. (b) Aerial view of the same scarp looking toward the northwest. (c) Close-up view of geologists and blocks on the upper portion of the scarp.

## 2.2. Rock Trait Extraction

Obtaining rock trait distributions from conventional methods presents challenges in human labor resources, scalability, and technology limitations. Laboratory analysis of field samples using traditional granulometric methods is time-consuming (Bunte & Abt, 2001; McCalpin et al., 1993). By leveraging image processing techniques, optical granulometry has advanced the field by using scale-calibrated digital photos to measure grain size distributions. Early optical granulometry adopted a two-stage process (Butler et al., 2001; Graham et al., 2005; Detert & Weitbrecht, 2012): the first stage binarized images by thresholding methods and the second stage applied watershed segmentation to generate grain instances.

Recent optical granulometry has employed machine learning techniques to segment rocks (Buscombe, 2013, 2020; Purinton & Bookhagen, 2019; Yang et al., 2021). However, the optical granulometry methods are constrained by discrete photo samples. That is, rock trait statistics such as mean, median, and standard deviation are extracted from a photo and then compared with statistics from other photos. These photos are discretely sampled from different locations, which cannot be directly applied to scenarios where spatially continuous distributions are of interest. Additionally, those methods require scale calibration for each photo to convert pixel counts to actual metric rock sizes.

Recent unpiloted aircraft system (UAS) and computer vision developments have facilitated geospatial image processing, enabling large-scale geostatistical analysis of rock traits (Carbonneau et al., 2018; Langhammer et al., 2017; Lang et al., 2021; Soloy et al., 2020). The methods in these studies generally involve two steps. First, from UAS imagery and ground control points, structure from motion (SfM) photogrammetry (Snavely et al., 2010) is used to produce georeferenced geometric models, such as point clouds, orthorectified maps (orthomaps), and digital elevation models (DEMs). Then, using computer vision techniques, rock trait distributions are estimated from sliced tiles of those georeferenced geometric models (usually orthomaps). The first step is commonly implemented with commercial tools (e.g., Agisoft Metashape); recent studies focused on the second step with different techniques to extract rock trait distributions. For example, BaseGrain (Detert & Weitbrecht, 2012), an advanced variant of the two-stage optical granulometry introduced above (Butler et al., 2001), was applied to UAS-SfM orthomaps for riverbed grain studies (Carbonneau et al., 2018; Langhammer et al., 2017). Lang et al. (2021) developed a deep learning regression model to directly estimate grain size statistics without segmenting individual rocks. Another recent study applied a novel deep neural network, Mask R-CNN (He et al., 2017), to segment individual rocks on UAS-SfM orthomaps for a grain size study of pebble beaches (Soloy et al., 2020). Applying those methods to orthomaps is advantageous because the rock trait distributions are georeferenced. However, because the estimated rock trait distribution is based on individual small tiles (8 x 8 meters in our case) cut from orthomaps, a rock on the edge of orthomap tiles may be split into two objects, resulting in an incorrect rock size estimation. For rocky fault scarps, splitting large rocks on the orthomap slice is especially problematic.

## 3. METHODS
### 3.1 Study Area
The Volcanic Tableland of eastern California (USA) is capped by the Bishop Tuff, a rhyolitic welded tuff produced at 758.9±1.8 ka by an eruption of the Long Valley Caldera volcano (Figure 3; van den Bogaard & Schirnick, 1995; Sarna-Wojcicki et al., 2000). The structurally controlled morphology of the welded ignimbrite and fault scarps that offset the relatively flat top of the Bishop Tuff, shown in Figure 2, are indicative of normal faulting processes (e.g., Dawers, et al., 1993; Ferrill et al., 2016; Pinter, 1995; Sheridan, 1970). We focused on a study area of 200 x 500 meters, annotated in Figure 3.

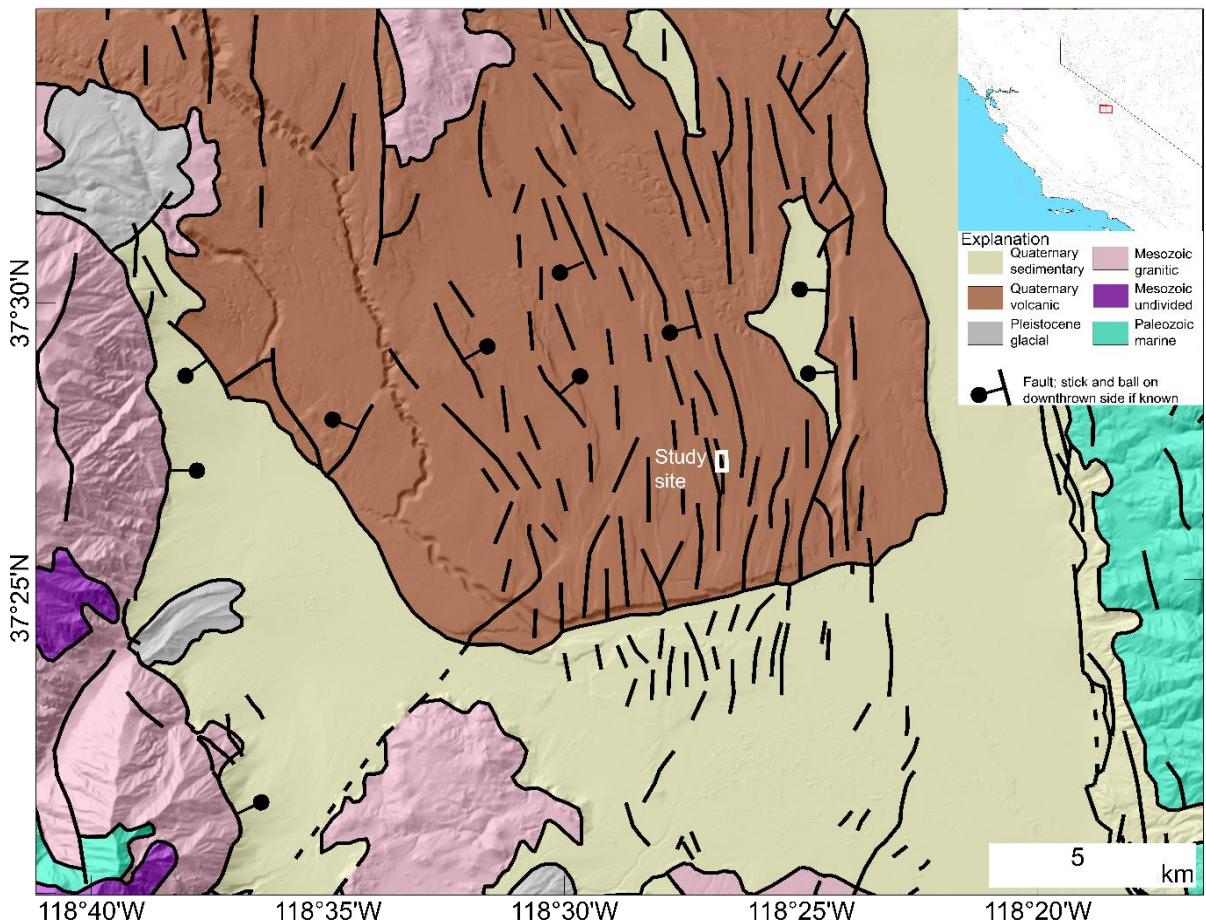

Figure 3. Area of study in eastern California (USA; after Ferrill, et al., 2016). The Volcanic Tableland (defined by the 760 ka Bishop Tuff) is indicated as the brown polygon which has been cut by faults (black lines) of the eastern California Shear Zone. Study area is shown as white rectangle. Geologic map modified from Jennings, et al., 2010. Inset at upper right shows active faults (USGS and CGS, 2021).

## 3.2 Data Acquisition

In March 2018, Scott et al. (2020) collected the UAS dataset that we used for our rock instance segmentation. They flew a DJI Phantom 4 Pro in lawnmower flight patterns with 66% image overlap and a 90° (nadir) camera angle, to survey a small portion of the Volcanic Tablelands. The flight altitude varied from 70 to 100 meters above the ground level, considering the maximum fault scarp height of 30 meters. The RGB camera on the DJI Phantom 4 Pro had an 84° field of view, 1-inch CMOS sensor, auto focus (at 1 meter and farther), and 5472 x 3648 pixel resolution. Agisoft Photoscan used the UAS survey images to reconstruct the site, from which we obtained an orthomap (2 cm/pixel) and DEM (4 cm/pixel) for deep learning. Scott et al. (2020) derived ground control from the image locations recorded by the DJI Phantom. Figure 4 shows the SfM products.

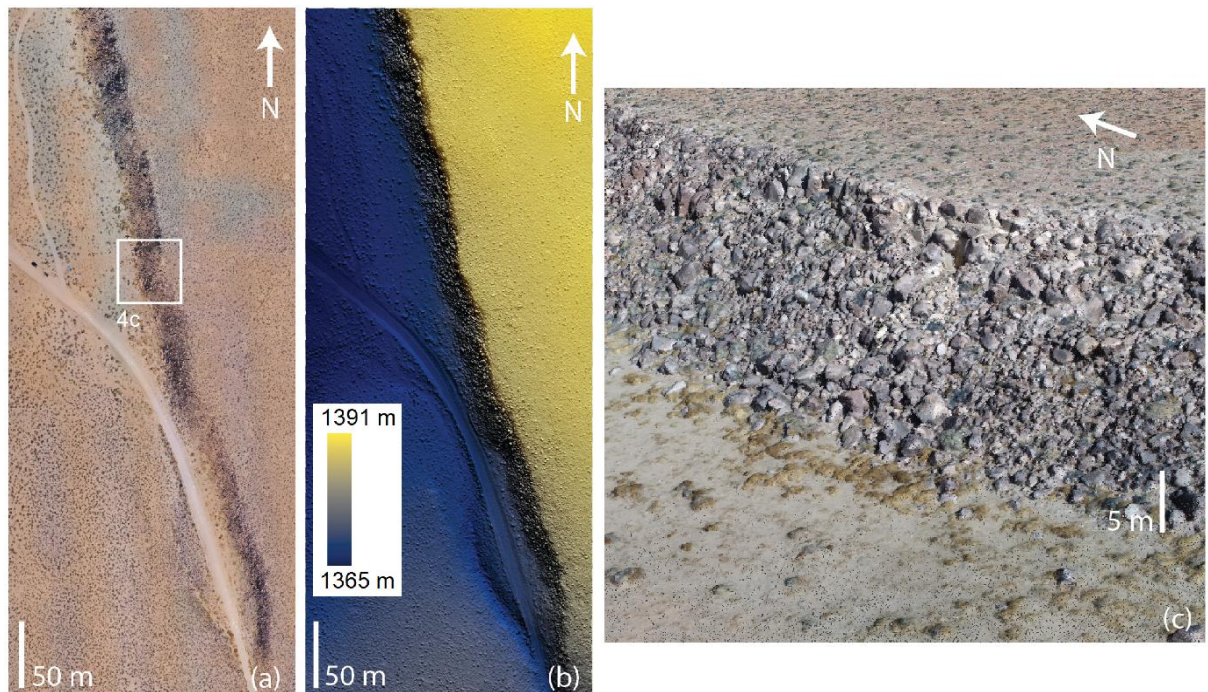

Figure 4. SfM products: (a) orthomap, (b) digital elevation model (DEM) with hillshade, and (c) 3D mesh model from rectangle area in orthomap. See Figure 3 for location.

For the correlation analyses with fault scarp geomorphic characteristics, we used fault scarp height from the recent work by Scott et al. (2022). They developed a semi-automatic algorithm to calculate fault scarp height from global and high-resolution topography datasets. Based on a manually drawn map of several square kilometers, the algorithm calibrated length and slope parameters to identify hanging wall and footwall flats as well as fault locations. The scarp height is the offset of the best-fit lines to the hanging wall and footwall flats at the scarp locations. To determine the scarp height, Scott et al. (2022) used a 0.5-meter resolution digital surface model produced from panchromatic tri-stereo images processed with Micmac software (Rupnik et al., 2016, 2017). The local fault scarp heights were calculated every 5 meters from 2-meter wide transects of the topography data.

We focused on a single fault scarp for this study. We segmented the scarp by applying a threshold to a DEM slope map (Appendix Figure A). With the segmented fault scarp, we removed outlier rocks that are not on the fault scarp, such that we could focus our analyses on the relationship between rock traits and fault scarp geomorphology. Additionally, on the segmented fault scarp, we conducted skeleton analysis to obtain the fault scarp middle spline (Zhang & Suen, 1984; see Appendix Figure A), which is close to the average fault trace. The skeleton analysis (Zhang & Suen, 1984) made successive passes to remove pixels on object borders until no more pixels could be removed, resulting in a fault scarp middle spline (Appendix Figure A).

## 3.3 Deep Learning

We employed Mask R-CNN (He et al., 2017) to segment individual rocks from the orthomap and DEM. Mask R-CNN is a deep learning method of instance segmentation, which segments and identifies each object of interest in an image. We selected Pyramid Feature Network (PFN; Lin et al., 2017) as the backbone of the Mask R-CNN. PFN's multi-scale, pyramidal-hierarchical anchor generation mechanism facilitated object detection for rocks with various pixel sizes in the images. The hyperparameter maximum number of regions of interest controlled the maximum number of rock candidates in an image. For our rocky fault scarp study, this hyperparameter was set to 256, which should be smaller than the largest number of rocks in an image. In our case, the largest number of rocks in an image is 225. For the benefit of repeatability and the interest of other potential applications, our code is open-source and available on GitHub: https://github.com/ZhiangChen/rock_traits

Appending depth information to RGB orthomaps improves instance segmentation performance (Chen et al., 2020). We applied the same input data strategy with four input channels: red, green, blue, and relative depth. The red, green, and blue channels were taken directly from the orthomap; the relative depth channel was calculated from the DEM. To get the relative depth channel, we first converted the DEM to a three-channel colormap (e.g., OpenCV colormap jet) where the elevations were represented by a set of RGB values. The relative depth map (RDM) is the average of the colormap channels:

$$D_{relative} = \frac{D_r}{3} + \frac{D_g}{3} + \frac{D_b}{3} \qquad (1)$$

where $D_r$, $D_g$, and $D_b$ are the RGB values in the DEM colormap.

To solve the rock splitting problem on image edges, we applied a scheme of overlap tiling and instance registration algorithm (Chen et al., 2020). Because the RDM had a lower pixel resolution (4 cm/pixel) than the orthomap (2 cm/pixel), we conducted bilinear interpolation to resample the RDM to the same size as the orthomap. Both the orthomap and RDM were split into 400 x 400 pixel tiles with 10-pixel overlaps on four edges. With new inference results from an input image, segmented rocks not in the four overlapped edges (e.g., rocks inside the square of overlap splitting lines in Appendix Figure B) were directly registered as new instances. Segmented rocks on the overlapped areas (four edges) were checked for their pixel overlaps with segmented rocks on the overlapped areas from the adjacent image tiles. They would be registered as new instances if their pixel overlaps were smaller than a threshold of 20; otherwise, they would be merged with the overlapped rocks from the adjacent image tiles. A result of the overlap tiling and instance registration is shown in Appendix Figure B. More details of the overlap tiling and instance registration can be found in Chen et al. (2020).

To improve the Mask R-CNN generalization, we selected 66 tiles from various lighting conditions, rock concentrations, and rock sizes. The selection depended on qualitative considerations such as visual cues, because the quantitative rock trait distributions were unknown before annotating. We annotated 66 tiles (with 4094 rocks). 49 tiles were used for training and 17 reserved for validation. The Mask R-CNN, initialized with well-trained weights from COCO 2017 (Lin et al., 2021), was re-trained on a server with an Nvidia RTX 2080 Ti Graphics Processing Unit. We selected the Mask R-CNN's weights from the best performance on the validation dataset to conduct inference on the tiles from the entire study area. Because the Mask R-CNN sometimes generated false bounding boxes smaller than 1 pixel, we removed such false-positive detection by keeping rocks with size greater than or equal to 2-cm diameter (1 pixel size).

We conducted the evaluation of the Mask R-CNN regarding the metrics from deep learning (instance segmentation) and rock trait statistics. On the validation dataset, we used mean Average Precision (mAP) and mean Average Recall (mAR) with various intersection-over-union (IoU) thresholds to measure the performance of object detection and segmentation (Chen et al., 2021). Because rock trait statistics were of interest in this study, we calculated the differences in rock trait statistics (mean and median of grain size) from neural network prediction and ground-truth measurements from annotation.

### 3.4 Semantic Rock Map
We have employed deep learning post-processing techniques to compensate for deep learning prediction errors and to extract rock traits. The segmentation sub-network in the Mask R-CNN sometimes generated stochastic errors in its output masks. As shown in Appendix Figure C, a hole was produced inside the segmented rock mask. Because rocks do not have torus topology in general, we applied topological structure analysis to obtain the contours of the segmented rock masks (Suzuki & be, 1985). From that, we retained the rock outlines from the largest or most exterior contours and removed the other contours. The rock outline resolution (polygon length) depended on the segmented rock mask resolution, which was 2 cm in this study. Further, we fit the resulting outlines with ellipses (Appendix Figure C), and approximated the aspect ratios and long-axis orientations of rocks with eccentricities and major axes of ellipses, respectively. The eccentricity of an ellipse is calculated by the formula,

$$e = \sqrt{1 - \frac{a^2}{b^2}} \qquad\qquad (2)$$

where $a$ is the length of the semi-major axis and $b$ is the length of the semi-minor axis. Eccentricities measure the roundness of an ellipse. Eccentricity of 0 indicates a circle.

To improve the efficiency of the geostatistical analysis, we built a geospatial database of the segmented rocks, presenting a semantic rock map. By leveraging the georeference of the orthomap, we generated a shapefile to represent the segmented rock outlines with georeferenced vector polygon objects. The shapefile format provides access to the segmented rock data via many geographic information system tools (e.g., ArcGIS and QGIS). The shapefile format also allows users to modify an individual rock if the instance segmentation is not ideal. In this work, we did not modify any neural network prediction because our evaluation shows low statistical error in the validation dataset. Additionally, we built a geospatial database using GeoPandas (Jordahl et al., 2020). Compared with brute-force 2D searching with quadratic time complexity of $O(N^2)$, the built-in functions in GeoPandas reduce the time complexity to linearithmic $O(2Nlog_2(N))$. The Big $O$ notation for time complexity describes the time to execute an algorithm in the worst-case scenario. For this study, the data searching by the linearithmic algorithm was noticeably more efficient than a quadratic algorithm because of the large instance number ($N$=230,310).

### 3.5 Data Analysis

Based on the semantic rock map (shapefile with rock-outline polygons; Figure 5), our data analysis focused on rock trait statistics and their correlations with fault scarp geomorphology. The polygons in the semantic map are 2D horizontal representations of the rock outlines. We acquired rock size distributions on the scarp and the surrounding topographic flats. To analyze the correlation between rock traits and fault scarp geomorphology, we calculated geospatial statistics only on the segmented fault scarp and not the surrounding topographic flats.

We quantified rock size distribution on the fault scarp and surrounding topographic flats over the entire semantic rock map. We sliced the entire area into a square mesh with a cell size of 5 meters. We calculated two rock size statistics in each cell: median effective diameter $D_{50}$ and median grain size $\Phi_{50}$:

$$D_{50} = median(D) \tag{3}$$
$$\Phi_{50} = median(\Phi) \tag{4}$$

where $D_{50}$ represents the 50-th percentile (median) rock effective diameter in millimeters, $D$ is an ordered list of rock effective diameters in the sample of interest, $\Phi_{50}$ represents the 50-th percentile (median) grain size, and $\Phi$ is an ordered list of grain sizes in the sample of interest. We calculate the effective diameter of a rock,

$$D = 2\sqrt{A/\pi} \tag{5}$$

where $D$ is rock effective diameter (in millimeters), $A$ is the area of a disk with the equivalent rock horizontal area. The rock horizontal area is determined from the polygon boundary of each particle. Grain size $\Phi$ is calculated from rock effective diameter (Krumbein, 1934):

$$\Phi = -log_2(D) \tag{6}$$

Note that, because of the negative sign in the formula, effective diameter decreases as the value of $\Phi$ increases. Appendix Table A presents a list of effective diameters and corresponding grain sizes $\Phi$. We used two rock size metrics because effective diameter is intuitive and grain size is a standard in granulometry (Krumbein, 1934). An advantage of using the logarithm conversion in grain size is to demonstrate the characteristics of small rocks in the grain size distribution (e.g., comparing Figure 6a, e and b, f). We constructed median rock size $D_{50}$ and median grain size $\Phi_{50}$ heatmaps (Figure 7). To analyze changes in rock size perpendicular to fault scarp trace, we horizontally shifted all rows in the mesh to align the fault scarp middle spline with the graph's long-axis (Figure 8). We focused on an analysis of the segmented rocks in the rectangular patch around the centerline trace. We sliced the patch into north-south lines with 5-meter spacing. In each column, we obtain a color-coded histogram of $\Phi_{50}$, as shown in Figure 8d. By stitching those color-coded histograms together, we produced a heatmap demonstrating changes in rock size perpendicular to the fault trace (Figure 8b).

On the segmented fault scarp, we studied the correlation between rock size and the local fault scarp height. We obtained 16 scarp sections by slicing the segmented fault scarp with east-west lines at 30-meter spacing (Appendix Figure A). Because the local fault scarp heights were calculated every 5 meters, we approximate each section's fault scarp height with the height of the nearest geolocation to the middle point of that section. We extracted $\Phi_{50}$ for each section. Using least squares linear regression, we calculated the r-squared value ($R^2$) and p-value of the correlation between the local scarp heights and $\Phi_{50}$. Additionally, because the large rocks exposed on the surface are an indicator of tectonic faulting, we analyzed the correlation between the local scarp height and the mean grain size of the largest ten rocks ($\overline{\Phi}_{M=10}$) from each fault scarp section:

$$\overline{\Phi}_M = mean(\Phi_{0:M}) \tag{7}$$

where $\Phi_{0:M}$ denotes a list of the smallest $M$ grain size $\Phi$ values in the sample of interest. Because grain size $\Phi$ and effective diameter $D$ are inverse, $\Phi_{0:M}$ also represents a list of the largest $M$ effective diameters in the sample of interest. To highlight the correlation between the local fault scarp height and small rocks, we inspected the ratio of small rock to large rock count, where the grain size $\Phi$ of -8 was the threshold to distinguish between small and large rocks. The threshold of -8 represents an effective diameter (25.6 cm) slightly larger than the median effective diameter. Besides small and large rocks, we also inspected the correlation between the local fault scarp height and sorting (or standard deviation, Folk and Ward 1957):

$$\sigma_\Phi = (\Phi_{84} - \Phi_{16}/4) + (\Phi_{95} - \Phi_5/6.6) \qquad (8)$$

where $\Phi_i$ denotes the $i$-th percentile grain size in the sample of interest.

We also investigated a hypothesis that a rock rolling downhill along a higher fault scarp has a greater likelihood of resting with its long axis parallel to the fault trace relative to a shorter scarp. To analyze the correlation between rock orientation and local fault scarp height, we sliced the segmented fault scarp into 16 sections at 30-meter spacing using the same process described above. Because the rock orientation statistics (e.g., long axis orientation) are potentially affected by the local fault trace, we represented a rock's orientation by a dot product between the ellipse's major axis direction and local fault trace. When a dot product result is greater than 0.5, the angle between the ellipse's major axis and fault trace is smaller than 45°, and the rock is a *tangent rock*. When the angle is greater than or equal to 45°, the rock is a *normal rock*. We used the ratio of the number of tangent rocks to normal rocks to indicate the degree of scarp-parallel rock orientation. A ratio that exceeds 1 indicates that there are more tangent rocks than normal rocks. We calculated this ratio from all rocks in each section and on the western half (or depositional portion) of the same section. We expected the rocks along the depositional portion of the scarp to have a stronger positive correlation with fault scarp height because the rocks below the middle spline have often been transported further downslope than those above.

## 4. RESULTS
### 4.1 Neural Network Evaluation Results
We evaluated the Mask R-CNN prediction performance by both instance segmentation metrics (Table 1) and grain size statistics (Table 2). As presented in Table 1, the instance segmentation showed good performance. For reference (He et al., 2017), the results of a similar neural network architecture trained from a large, comprehensive training data had slightly better performance on mAP(IoU=0.5) and mAR(IoU=0.5) of 51.2% and 55.2%, respectively. For object detection, the mAP and mAR of a list of IoU [0.5: 0.95] achieved 23.5% and 32.6%, respectively. The Mask R-CNN also showed good performance on segmentation with the mAP(IoU [0.5: 0.95]) and mAR(IoU [0.5: 0.95]) values of 21.6% and 30.7%, respectively. Additionally, the statistics from the prediction objects evaluated well with an absolute error for the median grain size of 0.01, with an equivalent effective diameter of 2.8 mm. We present samples of instance segmentation inference results in Figure 5.

Table 1. Instance segmentation evaluation on validation dataset. (mAP: mean average precision; mAR: mean average recall; IoU: intersection over union)

|  | mAP (IoU=[0.5:0.9]) | mAP (IoU=0.5) | mAP (IoU=0.75) | mAR (IoU=[0.5:0.9]) |
|---|---|---|---|---|
| Detection (Bounding Box) | 0.235 | 0.480 | 0.208 | 0.326 |
| Segmentation (Mask) | 0.216 | 0.484 | 0.165 | 0.307 |

Table 2. Grain size statistics evaluation on validation dataset.

|  | Mean effective diameter (mm) | Median effective diameter (mm) | Mean grain size ($\Phi$) | Median grain size ($\Phi$) |
|---|---|---|---|---|
| Test dataset | 390.7 | 401.7 | -8.61 | -8.65 |
| Prediction | 421.7 | 398.9 | -8.72 | -8.64 |
| Error | 31.0 | 2.8 | 0.11 | 0.01 |

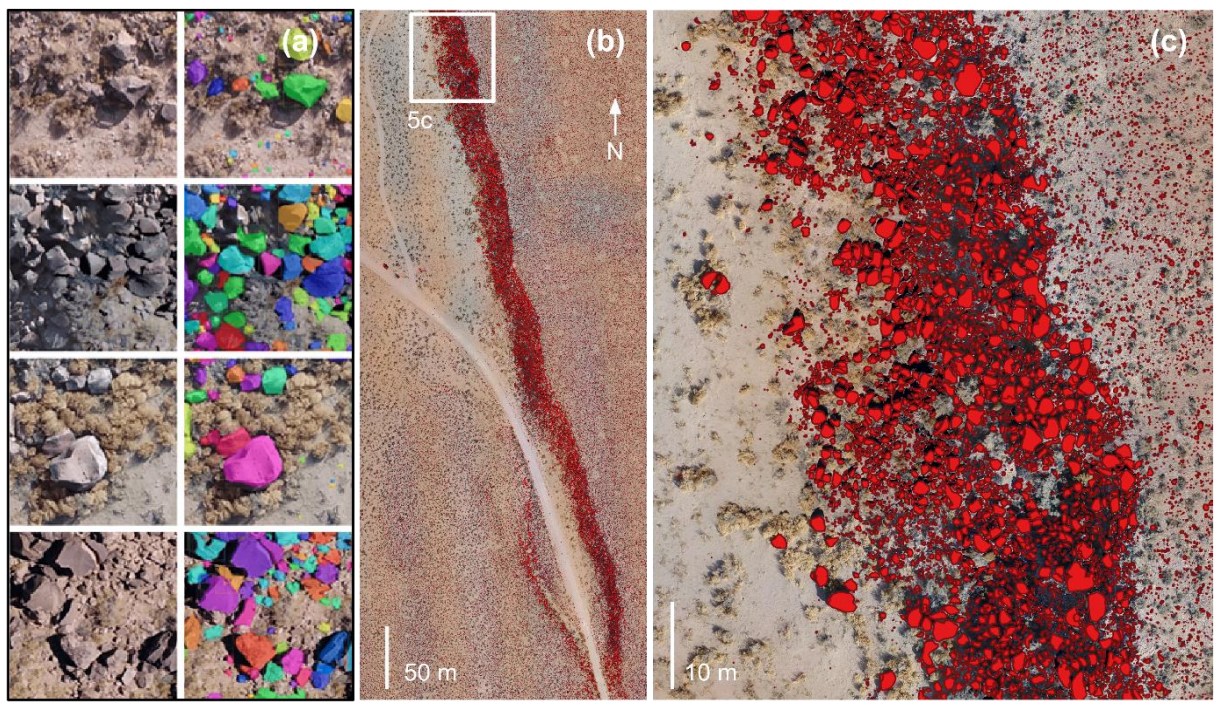

Figure 5. Instance segmentation inference results. In inference images (a, right column), colors distinguish individual rocks in input images (a, left column) from validation dataset. Rock polygons following instance registration are displayed in the (b) entire study area and (c) enlarged area.

## 4.2 Rock Trait Distributions

The Mask R-CNN detected 230,310 rocks on the fault scarp and surrounding flats including 36,133 rocks on the fault scarp. Figure 6 shows the rock trait histograms. On the scarp and surrounding flats, $\Phi_{50}$ is -6.22 (7.5 cm effective diameter), and 86.9% of rocks have eccentricities that exceed 0.5 (long and narrow rocks). On the fault scarp, $\Phi_{50}$ is -7.6 (nearly 18.8 cm effective diameter), and 92.8% of rocks have eccentricities greater than 0.5. Although both polar histograms (Figure 6) show ellipse-shaped distributions with peaks at the tangent and normal directions, the percentage of normal direction rocks on the fault scarp is higher than the one on the entire study area.

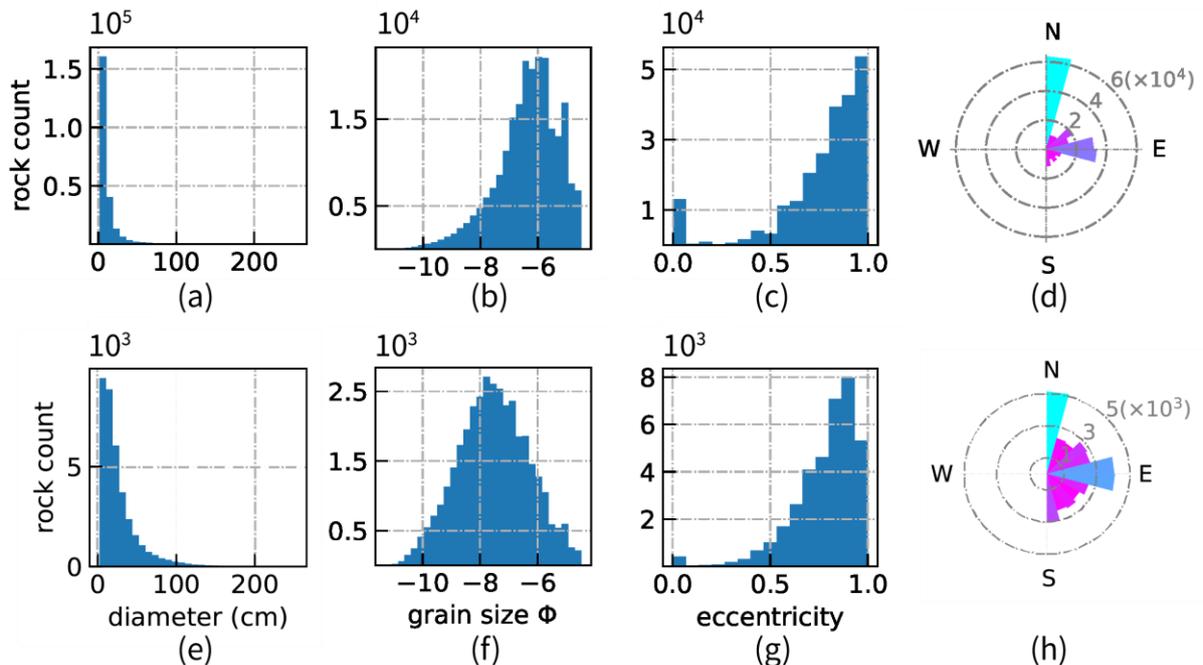

Figure 6. Effective diameter (cm), grain size, eccentricity, and long axis orientation histograms of all rocks (a, b, c, and d; on the fault scarp and surrounding topographic flats) and (e, f, g, and h) the fault scarp alone. Eccentricity of 0 indicates a circular rock. Polar histograms or rose diagrams show long axis orientation for a range of [0°, 180°) clockwise.

As shown in Figure 7, spatial distribution of the median rock sizes distinguishes the fault scarps from the topographic flats. Larger particles ($\Phi$ <-6; $D$ > 6.4 cm) are exposed on and below the fault scarps. The white pixels in the northwest of Figure 7b represent the absence of rock detection within the cells (rocks smaller than 2 cm diameter). Figure 8 depicts the median grain size change perpendicular to the fault scarp trace. From the hanging wall to the footwall (west to east), $\Phi_{50}$ gradually decreases from the base of the scarp where a range of particles have been deposited. Rocks with the largest effective diameters in general are at the upper portion of the scarp where the large columnar jointed cooling blocks on the footwall are exposed (Figure 2a).

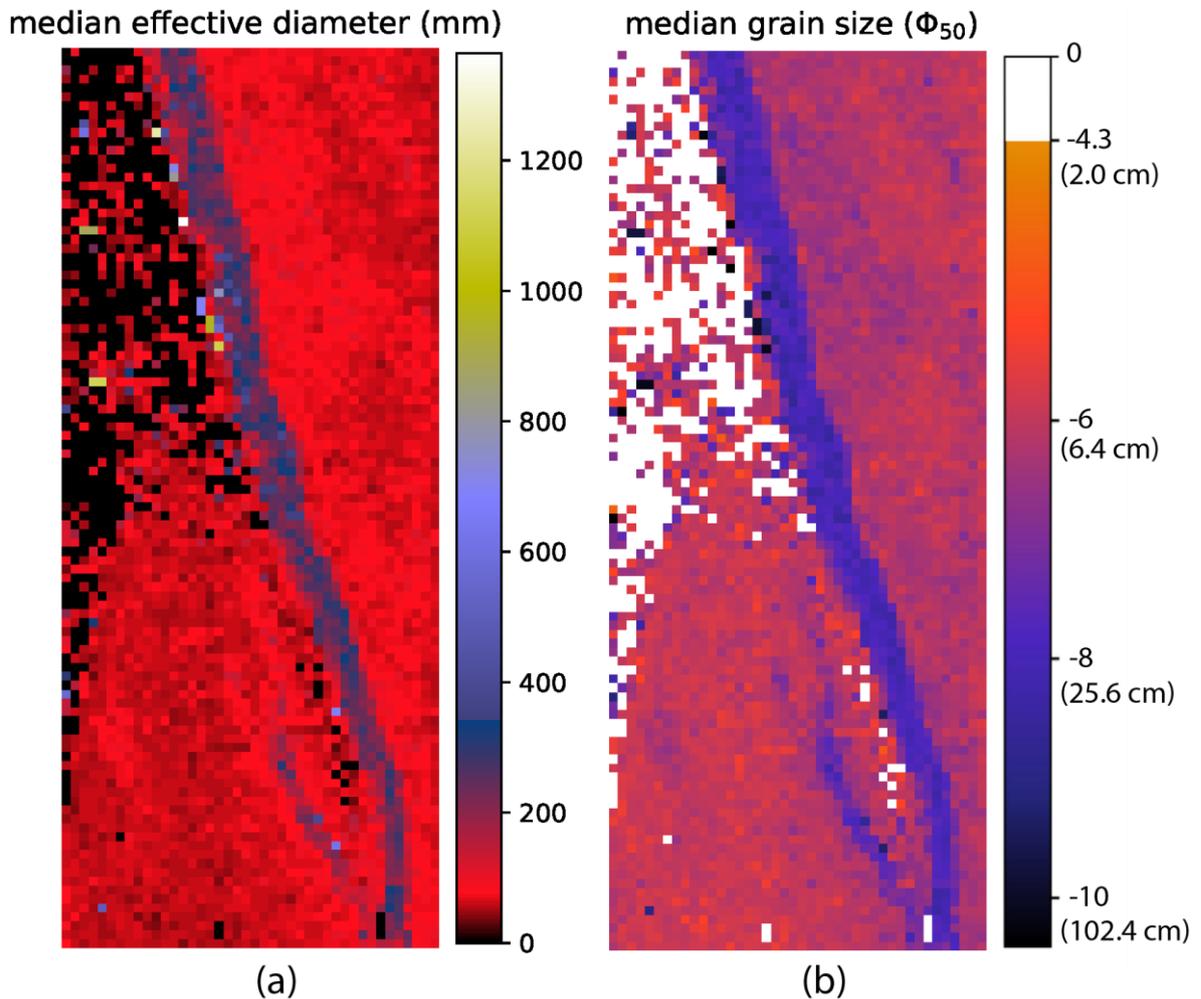

Figure 7. Spatial distributions of (a) median effective diameter and (b) median grain size ($\Phi_{50}$) over the entire study area. In the analysis, we sliced the entire study area into a square mesh with a cell size of 5 x 5 meters.

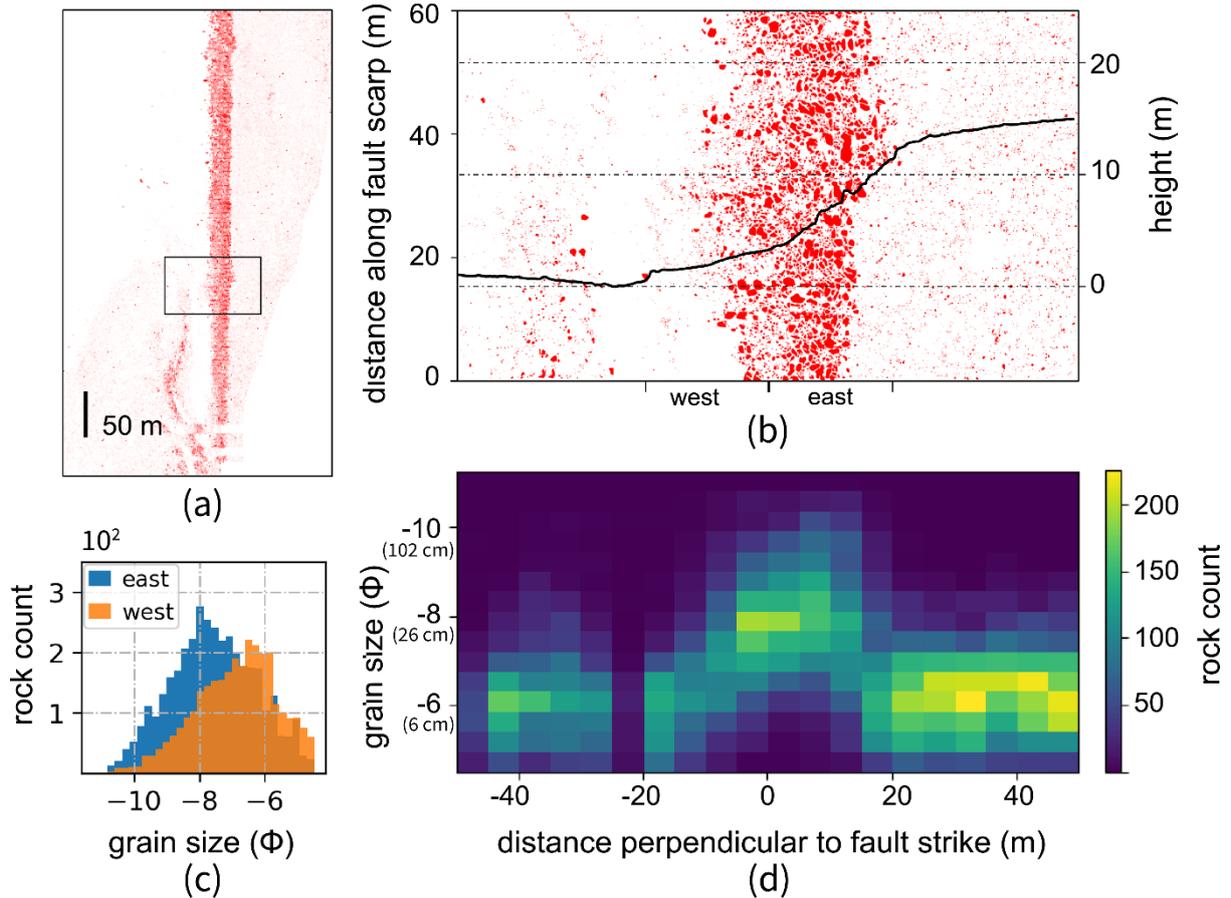

Figure 8. Spatial distributions of median grain size perpendicular to fault trace. (a) Rocks (red polygons) are horizontally shifted to align the fault scarp middle spline with the long axis of the image. (b) The scarp topographic profile (local height) horizontally sliced the inset at the middle of subfigure (a) is aligned with the background image. (c) Grain size histograms shows that median grain size from the western half ($\Phi_{50}$=-6.8, or 11.1 cm diameter) is greater than that on the eastern half ($\Phi_{50}$=-7.6, or 19.4 cm diameter). The western half is from -20 meters to 0, and the eastern half is from 0 to 20 meters, as annotated in subfigure (b). (d) Median grain size change perpendicular to the fault scarp trace is composed of color-coded histogram columns; the columns are sliced with 5-meter spacing; each color-coded histogram represents grain-size distribution in that column. Note that the grain size is plotted on a flipped positive to negative axis. Horizontal axes of subfigures (b, d) are aligned to show how the median grain size changes with the rock distribution and scarp topographic profile.

## 4.3 Correlation between Rock Traits and Fault Scarp Geomorphology

The local fault scarp height correlates strongly with the median grain size $\Phi_{50}$, the sorting $\sigma_{\Phi}$, the mean grain size of the largest ten rocks $\bar{\Phi}_{M=10}$, and the small to large rock count ratio, as shown in Figure 9 and Table 3. $\sigma_{\Phi}$ has the strongest correlation ($R^2$ of 0.81); the small to large rock count ratio has the lowest correlation ($R^2$ of 0.40). All the corresponding p-values are significantly smaller than 0.01 (Table 3), indicating strong evidence to reject the null hypothesis that the rock traits and fault scarp height are uncorrelated.

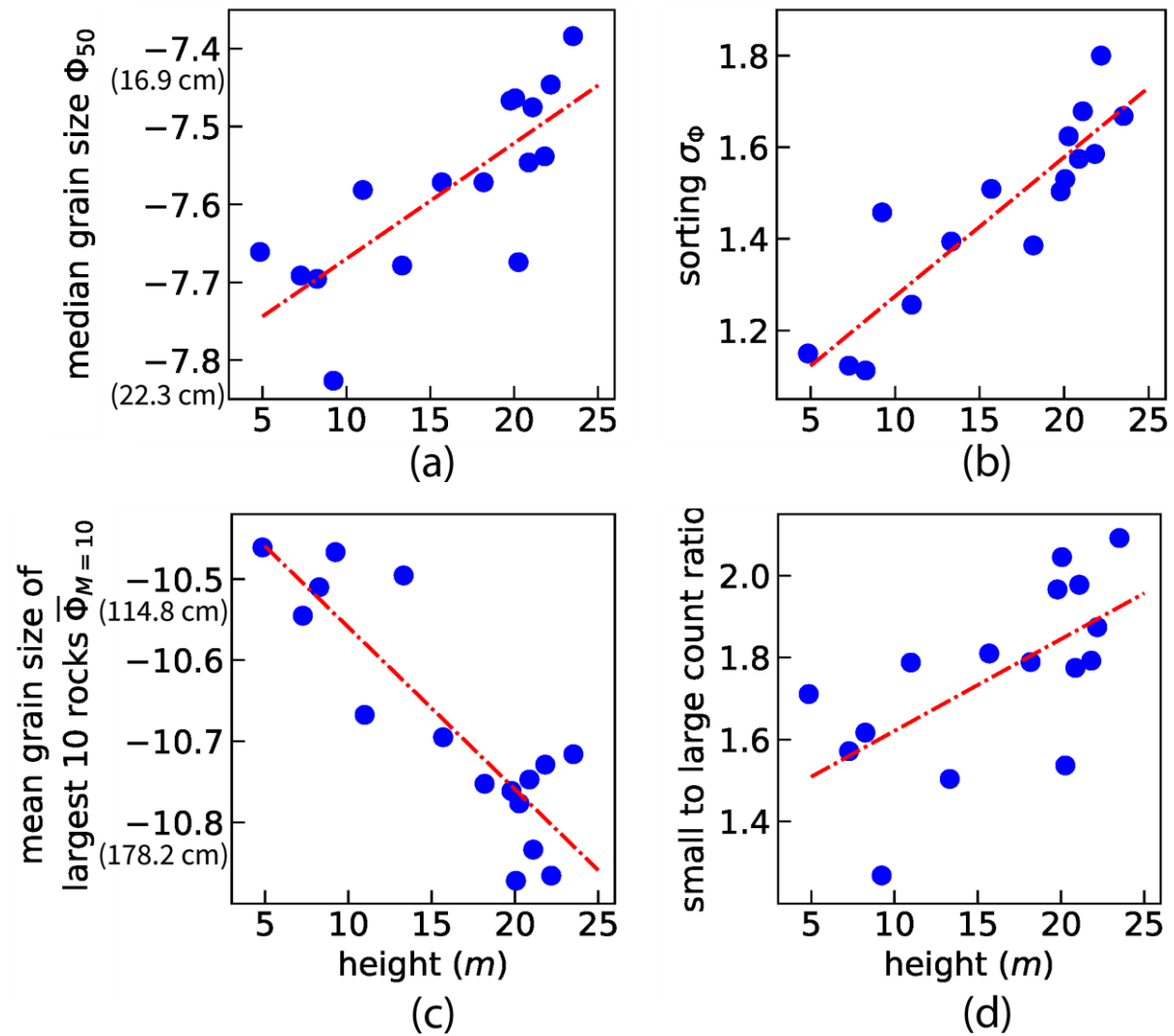

Figure 9. Correlation between local fault scarp height and (a) median grain size, (b) sorting, (c) mean grain size of the largest ten rocks, and (d) ratio of small rock to large rock count. Red dash-dotted plots are least squares linear regression lines.

In Figure 10a, b, inverted bell curves show rock orientation histograms for rocks from the entire fault scarp and the western half (or depositional portion) of the fault scarp. The two peaks in each histogram represent the rock long axes perpendicular and parallel to the fault scarp trace, respectively. The tangent to normal rock count ratio positively correlates with local fault scarp height, as shown in Figure 10c, d. Although the overall counts of tangent rocks and normal rocks are similar in Figure 10a, b, the tangent to normal rock count ratio from the western half of the segmented fault scarp correlates more strongly with the fault scarp height with the $R^2$ of 0.46, compared with the $R^2$ of 0.24 from the entire fault scarp (Table 3).

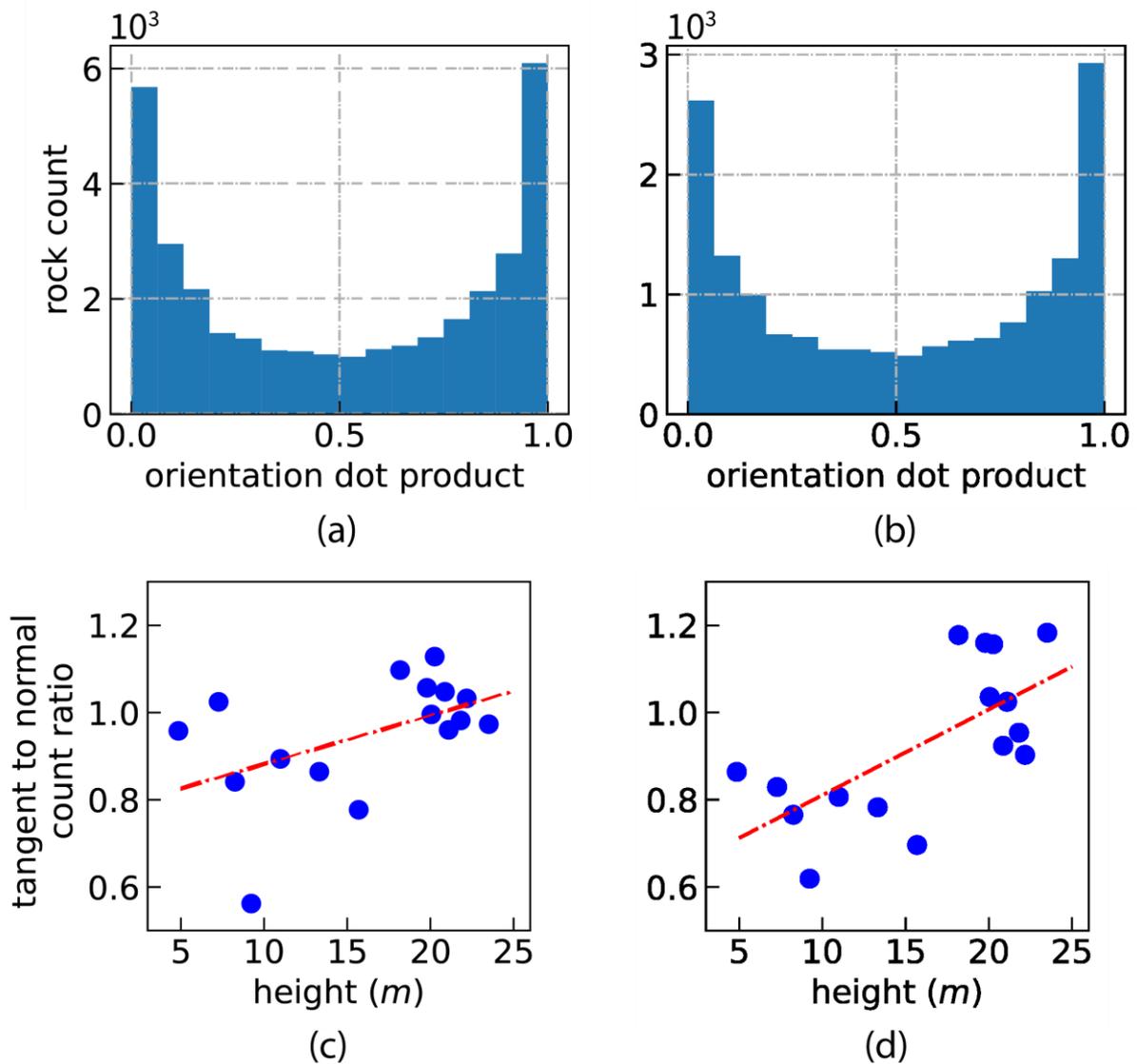

Figure 10. Rock orientation statistics and correlation with the local fault scarp height. (a, b) Rock orientation is represented by the dot product between the rock major-axis vector and local fault scarp trace. (c, d) Plots indicate the correlation between the local fault scarp height and the ratio of tangent rock count to normal rock count. Results came from (a, c) the fault scarp and (b, d) western half of the scarp.

Table 3. Rock traits and fault scarp height correlation statistics

| | $R^2$ | p-value |
|---|---|---|
| Median grain size $\Phi_{50}$ | 0.60 | $4.0 \times 10^{-4}$ |
| Largest grain size $\bar{\Phi}_{M=10}$ | 0.76 | $1.0 \times 10^{-5}$ |
| Sorting $\sigma_\Phi$ | 0.81 | $2.3 \times 10^{-6}$ |
| Small to large rock count ratio ($\Phi = -8$) | 0.40 | $8.5 \times 10^{-3}$ |
| Tangent to normal rock count ratio (whole scarp) | 0.24 | $5.3 \times 10^{-2}$ |
| Tangent to normal rock count ratio (western half) | 0.46 | $4.0 \times 10^{-3}$ |

## 5. DISCUSSION
### 5.1 Technology Improvements
We applied a Mask R-CNN to segment and identify individual rocks from an RGB orthomap and DEM that were derived from UAS aerial imagery of fault scarps in the Volcanic Tablelands, eastern California, collected in 2018. Although different lighting conditions caused stripes in the orthomap (Figure 4), the Mask R-CNN shows good inference results on the validation dataset according to the instance segmentation metrics and rock trait statistic metrics in Table 1 and Table 2. Geoscientists may collect UAS data in less than ideal conditions, because they need to cover a large area in a limited amount of time and thus cannot wait for ideal lighting. The Mask R-CNN inference results demonstrated that our annotation strategy and transfer learning techniques overcome the stripe artifacts from the various lighting conditions. To scale up to study areas where the SfM orthomaps are combined from multiple UAS surveys, more annotations from various lighting conditions and different areas are needed. Orthomaps with less distortion facilitate rock annotations and should also improve the instance segmentation performance for Mask R-CNN. To detect smaller rocks, a generic strategy is to increase orthomap resolution (cm/pixel) by altering the UAS flight plan (i.e., lower flight altitude) or SfM data processing configurations.

Distinguished from previous studies based on individual images (Soloy et al., 2020; Yang et al., 2021), our deep learning post-processing techniques enabled rock trait extraction and statistical analysis on a large scale (500-meter fault scarp with 2-cm resolution raster maps). The overlap tiling and instance registration solved the rock splitting problem where a single large rock may be split into two or more rocks. The rock contour analysis reduced false negative segmentation and helped to approximate aspect ratios as well as the orientation of the long axes of rocks. The semantic rock map enabled an efficient database management method of spatial data searching and expedited our statistical analysis of a large number of rocks.

Our big-data method of studying the correlation between rock traits and fault scarp geomorphology demonstrates two additional advances. First, eliminating machine learning errors is challenging. Regardless of the instance segmentation errors from the Mask R-CNN (Table 1), Table 2 shows the high accuracy of the rock trait statistics from a large number of segmented rocks. Second, because our training data were collected from the entire study area, the deep learning errors are consistent throughout all sections. Thus, the deep learning errors had a limited impact on the correlation analysis from different fault scarp sections. Future work should quantify the effects of deep learning errors on rock trait statistics, promoting the usage of rock trait statistics in broader geomorphology studies.

## 5.2 Grain Size Reduction and Grain Transportation on Rocky Scarps

We investigated rock trait distributions on the fault scarps and the surrounding topographic flats. We quantified rock size for several orders of magnitude more rocks than in previous studies (e.g., McCalpin et al., 1993; Ferrill et al., 2016), supporting statistical analysis of rock size. These data show that rock size distinguishes between the fault scarp and surrounding topographic flats (Figure 7). Our work serves as a powerful demonstration of the deep learning toolkit and has significant promise for large-scale, high-resolution examination of surface processes as indicated by rock traits and their distributions. Our method provides a new approach to detect and segment fault scarps.

We studied grain size change perpendicular to the fault scarp (Figure 8). The scarp stands out as increased effective diameters relative to the surrounding flats on either side. The top of the scarp has the largest effective diameter (presumably dominated by the original cooling columns with median effective diameters exceeding 1.0 m; Figures 3 and 8d). Moving down the scarp shows a decrease in median effective diameter with additional fracturing and particle transport.

Local fault scarp height correlates with rock size statistics (Figure 9 and Table 3), indicating that a locally higher fault scarp has smaller median effective diameter rocks as well as more small rocks relative to large rocks. Specifically, along the trace towards the north, the local fault scarp height tends to increase while the median effective diameter decreases, as shown in Figure 9a in terms of $\Phi_{50}$. The negative correlation between fault scarp height and the grain size of the largest ten blocks at any location, however, suggests that a locally higher fault scarp has larger rocks absolutely (Figure 9c). The positive correlation ($R^2$ of 0.40) between the local fault scarp height and the small to large rock count ratio further elucidates that a locally higher fault scarp has more small rocks relative to large rocks (Figure 9d). These correlation analysis results above are also consistent with the standard deviation of the grain size distribution being larger with increasing scarp height (Figure 9b): rocks on a locally higher fault scarp are less well sorted. Collectively, these correlations suggest that higher portions of the fault scarp have exposed larger blocks, and those portions of the scarp have smaller rocks and more small rocks that have been accumulated from fracturing and downslope transport. We cannot quantitatively reverse engineer the contributions of the different fracturing processes to the final grain size distribution.

We investigated the correlation between the local fault scarp height and rock orientation to study particle transportation. McCalpin et al. (1993) proposed a particle transportation model where a long and narrow rock rolling down a slope will have its long axis oriented parallel to local fault scarp trace. We extend this model to hypothesize that locally higher fault scarps will have relatively more tangent rocks because the rocks have a larger distance to roll and orient the long axis. This model was supported by our correlation analysis between the local fault scarp height and tangent to normal rock count ratio (Figure 10 and Table 3). Both the tangent to normal rock count ratios from the entire fault scarp and the western half of the fault scarp tend to increase with the local fault scarp height. Such correlations indicate that locally higher fault scarps have more tangent rocks. A higher ratio of tangent to normal rock count is expected for the western half of the segmented fault scarp, because rocks below the fault scarp middle spline have more distance to roll downslope. We analyzed the correlation between the local fault scarp height and the ratio of tangent to normal rock count, and observed that western half of the fault scarp has a stronger correlation ($R^2$ of 0.46) than the entire fault scale ($R^2$ of 0.24; Table 3), supporting this hypothesis.

### 5.3 Limitations and Future Work
Future work should examine the effects of obscured rocks on rock trait distribution. How Mask R-CNN detects partially obscured objects depends on annotations in training data. In our case, when a rock is partially obscured, we used a polygon to outline the visible portion of the rock, excluding the visually obscured part, as the annotation. We did this because how much the rock is obscured is unknown, and making the neural network conduct inference based on incomplete information may cause stochastic results. Additionally, for correlation analysis, we used the statistics of rock traits where the effects of obscured rocks were reduced because the fraction of obstructed rocks was small. However, quantifying the effects of obscured rocks on rock trait distributions and correlation analysis is important for future work.

Future technical work should be aimed at reducing uncertainties introduced by 2D rock representations of 3D rocks. Although we used the 2.5D data (RGB orthomaps and DEM) as the Mask R-CNN inputs, the rock traits (rock size, eccentricity, and orientation) were still extracted from the 2D masks of the neural network prediction. Future work should quantify the uncertainties of using 2D to represent 3D geometry. A more radical solution to reducing the representation uncertainties is to employ 3D point segmentation, such as KPConv (Thomas et al., 2019). Because of increased dimensionality, 3D point segmentation should yield better rock trait estimates, resulting in more accurate estimates of rock trait statistics.

The correlation analyses for geomorphological studies of rocky fault scarps should be further validated at larger scales across the Volcanic Tablelands and other rocky fault scarps in the future. This study indicates that rock trait statistics have a relationship with fault scarp formation processes. Larger-scale surveys and analyses will scrutinize this relationship. A larger parameter space of rock types, faulting history, and surface processes with additional field observations will help to better separate and quantify the relative roles of fracture processes on rock trait inventories. Large-scale surveys with different scarp height scales will increase understanding of the effects of changing rates of geomorphic and tectonic processes on rock trait distributions. Additionally, the distributions of rock traits might also be used as constraints in Probabilistic Seismic Hazard Analysis models (Youngs et al., 2003) to reduce uncertainties in ground motion estimation. Finally, these techniques could be applied to other geologic problems, such as grain characteristics of explosive volcanic eruptions, which are a critical part of quantifying eruption processes, and those on other planetary surfaces such as Mars where exposed columnar joints have been documented (Milazzo et al. 2009).

## 6. CONCLUSIONS
We adopted the UAS-SfM-DL approach to segment individual rocks on a fault scarp near Bishop California (USA) based on imagery collected in 2018. The deep learning showed good instance segmentation performance on the validation dataset with object detection mAP(IoU=0.5) of 0.480 and segmentation mAP(IoU=0.5) of 0.484. We examined the median grain size on the validation dataset with an absolute error of 0.01 (median effective diameter error of 2.8 mm). To solve the rock splitting problem on image edges, we applied a scheme of overlap tiling and instance registration, resulting in a semantic map of rocks. From the semantic map, we quantified spatial variations in rock size on the fault scarp and surrounding topographic flats. We visualized the results on two heatmaps. We also inspected how median grain size changes perpendicular to the fault scarp trace. To study the relationship between rock traits and fault scarp geomorphology, we conducted correlation analyses on the segmented fault scarp. The rock traits indicate that higher fault scarps have accumulated more geomorphic fracturing and tectonic faulting. Rock orientation statistics and local fault scarp heights provided evidence for a particle transportation model. While we demonstrated a data-driven approach to geomorphological analysis of rocky fault scarps, the UAS-SfM-DL pipeline has great potential to be used as a generic tool for other geomorphological analysis applications using rock trait distributions.

## ACKNOWLEDGEMENTS
This work was supported in part by Southern California Earthquake Center (SCEC) award 19179, National Science Foundation award CNS-1521617, NASA Flight Opportunities program, and Pacific Gas and Electric. Discussions with W. Ashley Griffith and Abdel Hafiz have helped us to refine our thinking on rocky scarps. Thank you to Adam Cawood, an anonymous reviewer, the associate editor, and Editor Stuart Lane for their efforts to improve the manuscript.

**APPENDIX**

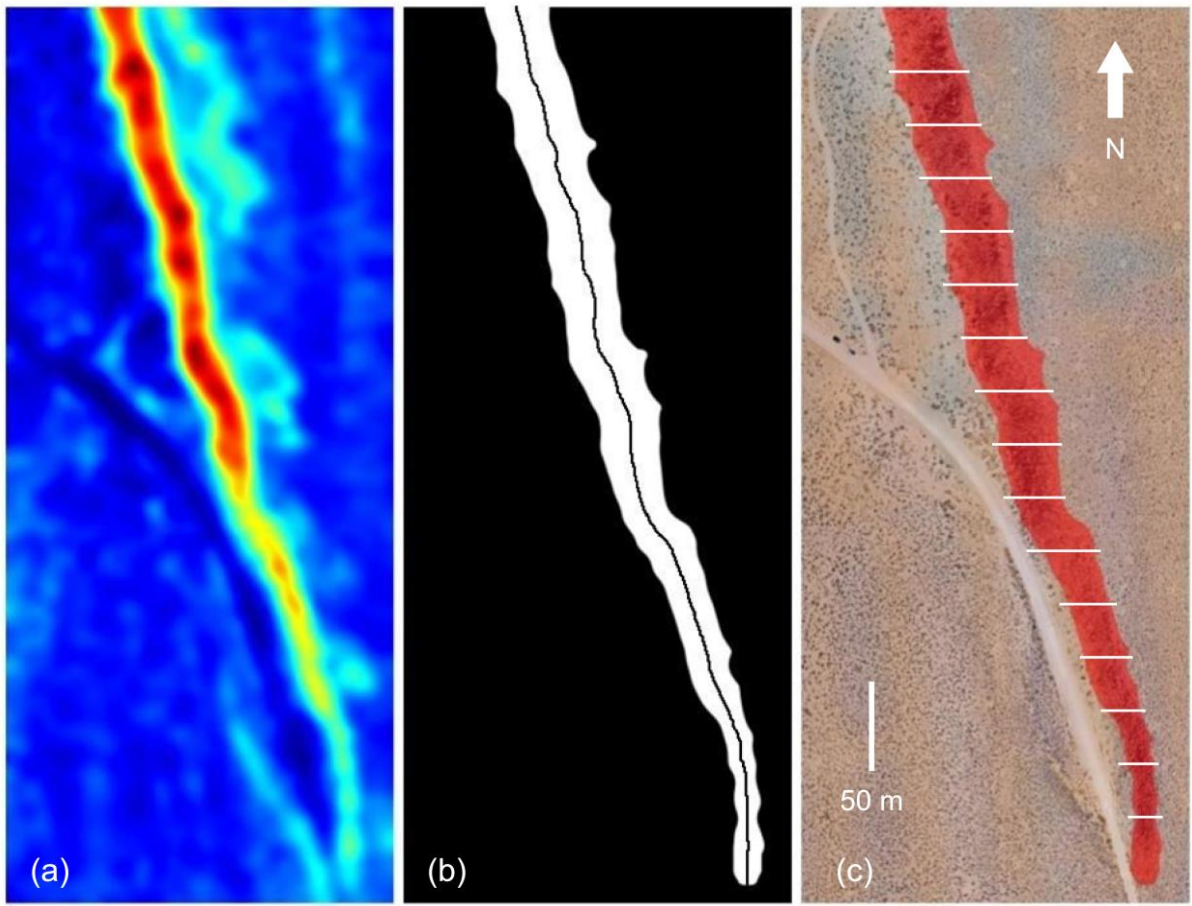

Figure A. (b) Segmented fault scarp generated by thresholding (a) slope map. (b) Middle spline of the segmented fault scarp is obtained from skeleton analysis. (c) Overlap visualization of segmented fault scarp and orthomap. The segmented fault scarp is sliced with east-west lines at 30-meter spacing. Figure is adapted from Chen et al., 2020.

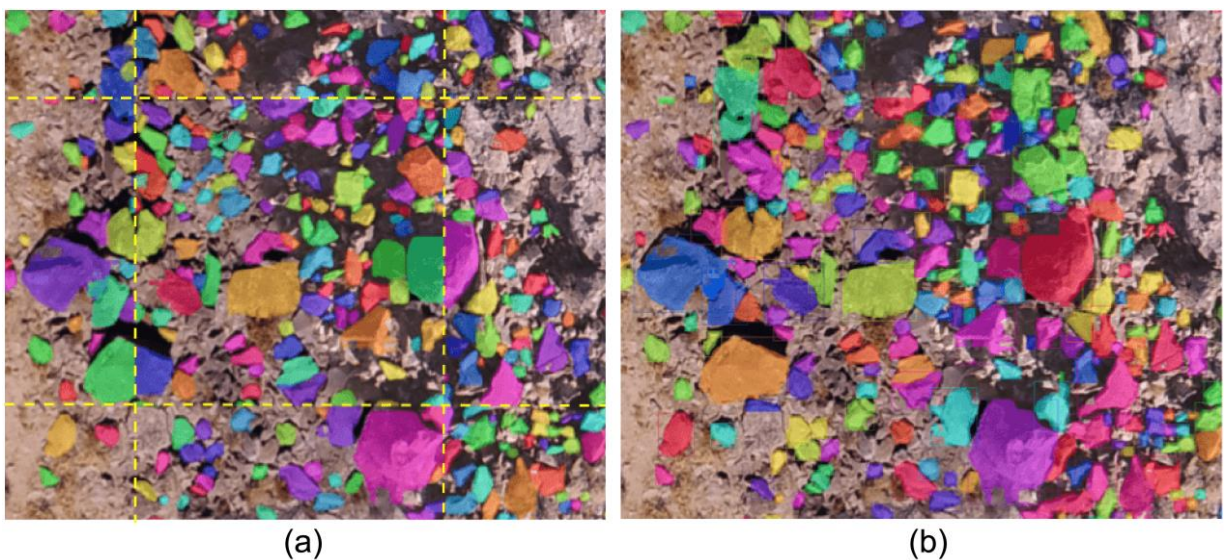

Figure B. Result of overlap tiling and instance registration. (Left) Rocks are split by tile boundaries, as opposed to (right) merged rocks via overlap tiling and instance registration strategy. Figure is adapted from Chen et al., 2020.

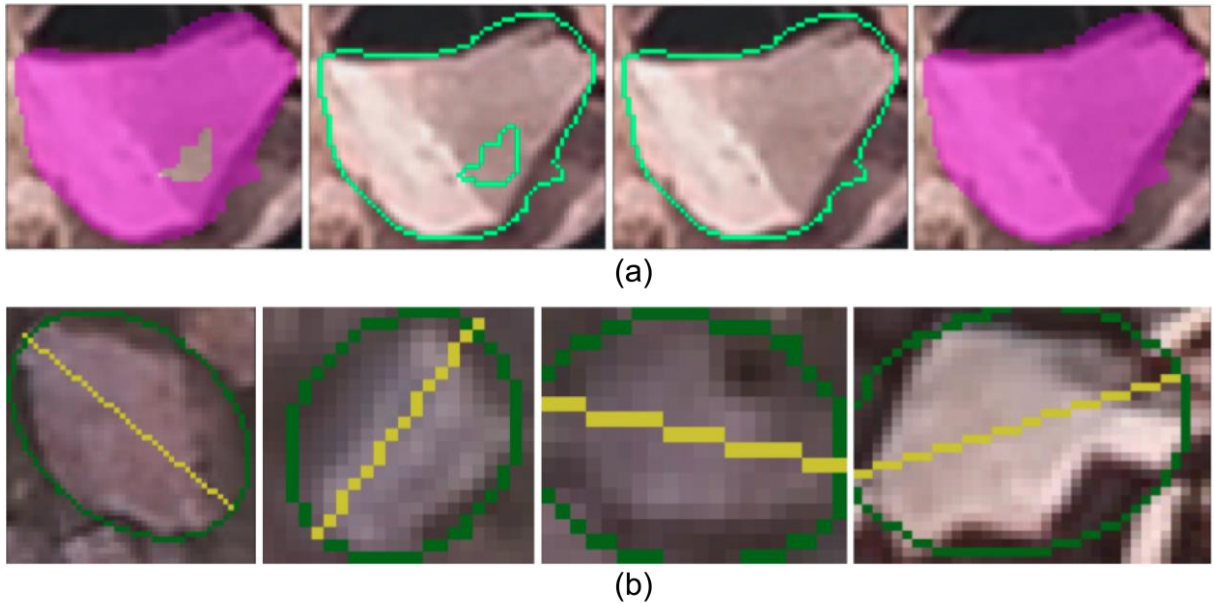

(a)

(b)

Figure C. (a) Rock contour analysis and (b) ellipse approximation. (a) Hole on segmentation mask is generated from Mask R-CNN inference and removed by contour analysis. (b) Rock orientations are represented by major axes of fitted ellipses (Fitzgibbon et al., 1999). Figure is adapted from Chen et al., 2020.

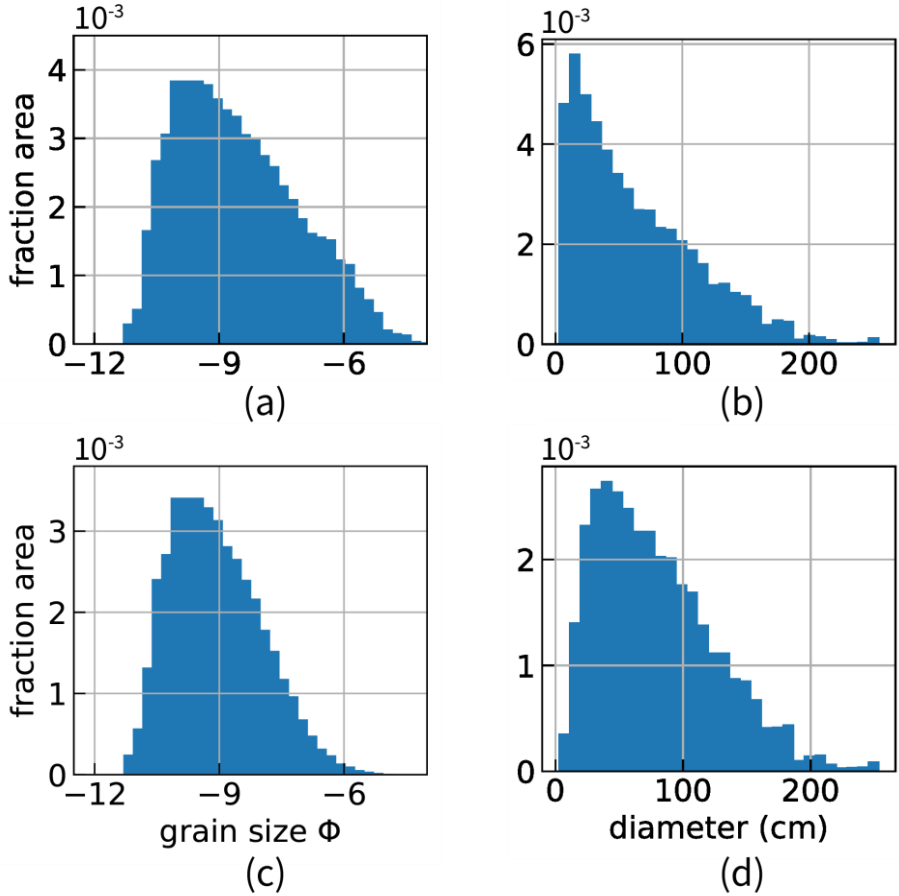

Figure D. (a, c) grain size and (b, d) effective diameter histograms on (a, b) the fault scarp and surrounding topographic flats and (c, d) the segmented fault scarp. Fraction area is calculated by $F_i = \frac{\Sigma\{A\}_i}{S}$, where $S$ is the entire study area including the fault scarp and surrounding flats, and $\Sigma\{A\}_i$ is the sum of rock areas in the $i$-th bin.

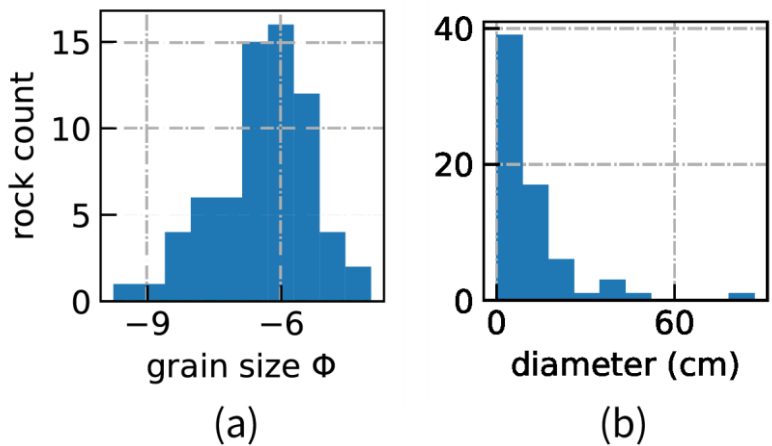

Figure E. (a) Grain size histogram and (b) effective diameter histogram in a 5x5 m cell obtained from importance sampling (a Monte Carlo sampling method) of the fault scarp and surrounding flats. The total sample size is the average rock count in a cell. Assuming all cells have equal importance (same sampling weights) and all rocks in a cell have equal importance, the data are sampled from each cell.

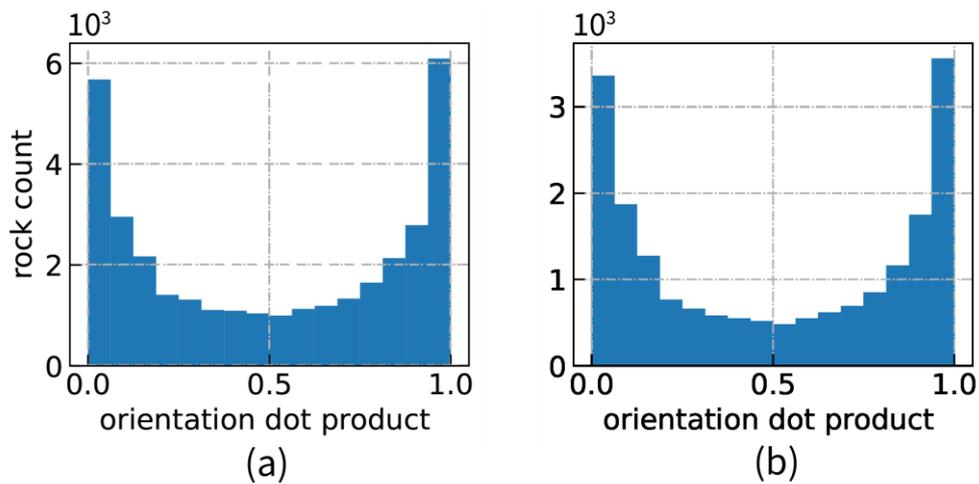

Figure F. Orientation histograms of (a) all rocks and (b) rocks with eccentricities larger than 0.8. These two histograms show the consistent saddle-shape pattern.

Table A. Mapping between effective diameter and corresponding grain size

| Effective diameter (mm) | Grain size $\Phi$ |
| --- | --- |
| 1 | 0 |
| 10 | -3.3 |
| 100 | -6.6 |
| 1000 | -10.0 |
| 5000 | -12.3 |